\documentstyle[12pt,aasms4,epsfig]{article}

\begin{document}

\title{Gamma-Ray Bursts and the Cosmic Star Formation Rate}
\author{Mark Krumholz\altaffilmark{1} and S. E. Thorsett\altaffilmark{2}}
\altaffiltext{1}{krumholz@pulsar.princeton.edu}
\altaffiltext{2}{Alfred P. Sloan Research Fellow; steve@pulsar.princeton.edu}
\affil{Joseph Henry Laboratories and Department of Physics, Princeton
  University, \\Princeton, NJ 08544}

\and

\author{Fiona A. Harrison\altaffilmark{3}}
\altaffiltext{3}{fiona@srl.caltech.edu}
\affil{Space Radiation Laboratory and Division of Physics, Mathematics
and Astronomy, California Institute of Technology, Pasadena, CA 91125}

\setcounter{footnote}{0}
%\date{\today}

\renewcommand{\baselinestretch}{1}
\small\normalsize

\begin{abstract}

We have tested several models of GRB luminosity and redshift
distribution functions for compatibility with the BATSE 4B number
versus peak flux relation. Our results disagree with recent claims
that current GRB observations can be used to strongly constrain the
cosmic star formation history. Instead, we find that relaxing the
assumption that GRBs are standard candles renders a very broad range
of models consistent with the BATSE number-flux relation. We
explicitly construct two sample distributions, one tracing the star
formation history and one with a constant comoving density.  We show
that both distributions are compatible with the observed fluxes and
redshifts of the bursts GRB970508, GRB971214, and GRB980703, and we
discuss the measurements required to distinguish the two models.

\end{abstract}


\section{Introduction}

The recent measurements of redshifted absorption lines in the optical
counterpart associated with GRB970508 (\cite{mdk+97}), and emission
lines from the galaxies associated with GRB971214 (\cite{kdr+98}) and
GRB980703 (\cite{dkg+98}) have confirmed the cosmological origin of gamma
ray bursts. Such measurements offer a new method of estimating the
distance and luminosity distribution of GRBs.  Determining the
redshift distribution, in particular, is of enormous importance in
distinguishing between GRB source models.  Much attention has focused
on models in which burst activity is tied to the massive star
formation rate, either directly (e.g., the ``failed supernova'' models
of Woosley 1993\nocite{woo93}), or with a time delay (e.g., the
in-spiraling neutron star model of Pacy\'nski 1986\nocite{pac86}). In
such models, GRB activity should trace the cosmic star formation
history (\cite{mad97}; \cite{fcp96}).

Previous analyses have attempted to constrain the GRB distribution
using the cumulative number versus peak flux, $N(<S)$, relation
observed by the BATSE instrument ({\em e.g.} \cite{rhl95}).  By making
additional assumptions about the GRB luminosity distribution (usually
that bursts are ``standard candles''), various authors have reached
strong conclusions about the GRB distribution.  For example, Totani
(1998)\nocite{tot98} recently used GRB observations to argue that our
understanding of the cosmic star formation rate is incorrect.

As we show in \S\ref{maps}, relaxing the constraints on the
intrinsic GRB luminosity function greatly reduces the power of GRB
number counts to constrain the burst distribution.  Standard candle
models are no longer tenable; there are now redshift measurements of
three bursts that vary in intrinsic luminosity by a factor of
20.\footnote{Unless otherwise stated, all models assume an $H_0=75$ km
$\mbox{s}^{-1}$ $\mbox{Mpc}^{-1}$, $\Omega_0=1.0$, $\Lambda=0$
cosmology, with no beaming.} We demonstrate that a wide range of burst
distributions are allowed by the current data, including both the
BATSE measurements and the measured burst redshifts.  However, we show
in \S\ref{case} that further redshift measurements, as well as deeper
number counts, can be used to ``deconvolve'' the unknown redshift and
luminosity distributions.

\section{Theory}

\subsection{Tests Using Number Versus Flux}

A source at redshift $z$ has an integrated comoving luminosity in the bandpass
$\left[\nu_1,\nu_2\right]$ at the source given by
$L=\int_{\nu_1}^{\nu_2}\frac{dL}{d\nu}d\nu$,
where $dL/d\nu$ is the spectrum of the source. If we assume a spectral form
$dL/d\nu=C\nu^{-\alpha}$, we can compute the
luminosity $L'$ at Earth in the bandpass
$\left[\nu'_1,\nu'_2\right]$, where 
$\nu'=\nu/(1+z)$, using
\begin{equation}
\label{lumdef}
L' = \frac{1}{1+z} \int_{\nu'_1}^{\nu'_2} C\nu^{-\alpha}d\nu'
 =  (1+z)^{-\alpha}\int_{\nu_1}^{\nu_2}C\nu^{-\alpha}d\nu
 =  (1+z)^{-\alpha}\cdot L.
\end{equation}
The additional factor of
$1+z$ in equation (\ref{lumdef}) comes from time dilation at the
burst source. The flux from the source observed at Earth is
(\cite{rhl95})
\begin{equation}
\label{zLflux}
S\left(z,L\right)=
	\frac{L \left(1+z\right)^{-\alpha}}{4\pi R_0^2 r(z)^2}.
\end{equation}
The coordinate distance, $r(z)$, is given by Weinberg (1972) \nocite{wei72}
\begin{equation}
r(z)=\frac{zq_0+(q_0-1)\left(\sqrt{2q_0z+1}-1\right)}{H_0 R_0 q_0^2 (1+z)}.
\end{equation}
Here $R_0$, $H_0$, and $q_0$ are the present-epoch expansion, Hubble,
and deceleration parameters.

A model for GRB sources consists of a distribution in comoving
redshift space, $dp/dz$, and a distribution in luminosity space,
$dg/dL$. The redshift distribution has three parts: a geometric
factor that accounts for the differential volume available at redshift
$z$, a factor to include relativistic effects, and a factor $n(z)$
for the comoving differential density of GRBs per unit volume at
redshift $z$. Thus, $dp/dz$ is given by
\begin{equation}
\frac{dp}{dz} =  \frac{dV}{dz}\frac{1}{1+z}n\left(z\right) 
  =  4\pi \frac{r(z)^2}{\sqrt{1-kr(z)^2}} \frac{dr(z)}{dz} \frac{n\left(z\right)}{1+z}.
\end{equation}
Here, $dV/dz$ is the comoving volume element per unit redshift, the factor
$1/(1+z)$ comes from time dilation at the burst source, and $k$ is $-1$, 0,
or 1 for open, flat, and closed cosmologies, respectively.

To test a model, we generate a large number of bursts. We select
redshifts from the distribution $dp/dz$, and luminosities from
$dg/dL$. Using equation (\ref{zLflux}), we determine the flux for
each burst in the sample, correct for flux-dependent instrumental
sensitivity, and form a cumulative distribution function $N(<S)$. We
compare the distribution function for the model to
the observed $N(<S)$ relation using a Kolmogorov-Smirnov (KS) test,
calculating a confidence level at which that model can be ruled out.

\subsection{Tests Using Redshift Measurements}

With redshift measurements, we can compare a model's predicted redshift
distribution to the observed distribution. It is necessary to correct
the predicted distribution for the bias introduced by a
finite flux cutoff. To do this, we generate a large number of bursts
from the model, correct the sample for instrumental sensitivity, and
simultaneously compare the redshift and flux distributions using a
two-sided, two-dimensional KS test (\cite{ptvf92}).

\section{Maps of Parameter Space}
\label{maps}

We examine range-dominated power law luminosity distributions, which have
previously been studied by Emslie \& Horack (1994)~\nocite{eh94b},
Hakkila {\em et al.} (1996)~\nocite{hmh+96}, Horack {\em et al.}
(1996)~\nocite{hhe+96}, Horv\'{a}th {\em et al.}
(1996)~\nocite{hmm96}, and Loredo \& Wasserman
(1998)~\nocite{lw98}. These distributions have
the form $dg/dL=L^p$ in the range $[L_{\rm min}, L_{\rm max}]$ and are
zero everywhere else. For the star formation rate, we use the
closed-box model of 280 nm cosmic emissivity, including dust
obscuration, taken from Fall {\em et al.} (1996). \nocite{fcp96}
This paper and others (e.g. \cite{mad97}) include several other
estimates of the cosmic star formation history. We choose this model
as an example; other SFR models would give numerically
different but qualitatively similar results.

First, we find the comoving luminosity of the three GRBs in the BATSE
bandpass (50-300 keV). GRB971214 has a peak flux of 1.9538 photons
$\mbox{cm}^{-2}$ $\mbox{s}^{-1}$ on the 1024 ms time-scale, and a
measured redshift of 3.42 (\cite{kdr+98}). GRB970508 has a peak flux
of 0.969 photons $\mbox{cm}^{-2}$ $\mbox{s}^{-1}$, absorption features
at $z=0.835$ (\cite{mdk+97}), and it most likely in a galaxy at that
distance (\cite{pfb+98}).  GRB980703 has a peak flux of 2.42 photons
$\mbox{cm}^{-2}$ $\mbox{s}^{-1}$, and is associated with an object at
$z=0.967$(\cite{dkg+98}).  We obtained the BATSE peak fluxes from
Meegan {\em et al.} (1998)~\nocite{mpb+98}.

To use equation (\ref{zLflux}), we need to know $\alpha$, the GRB
spectral index. GRB spectra vary considerably (\cite{bmf+93}); a
thorough treatment would assume a distribution of underlying spectra
in its analysis. However, as we will see, the current data do not yet
justify this expansion of the model phase space. Instead, we simplify
by using as our single spectral index the recent estimate $\alpha=1.1$
(\cite{mpp96}). With this assumption, we calculate luminosity
$L_1=2.15\times 10^{50}$ erg/s for GRB970508, 
$L_2=4.60\times 10^{51}$ erg/s for GRB971214, and
$L_3=6.70\times 10^{50}$ erg/s for GRB980703.

We examine the section of parameter space $L_1 /10 <
L_{\rm min}< L_1$, $L_2 < L_{\rm max}< L_2 \cdot 10$,
$-1<p<-2.5$, on a 21 by 21 by 11 grid evenly spaced in the logarithm of
$L_{\rm min}$ and $L_{\rm max}$, and evenly
spaced in $p$.

The BATSE GRB catalog (\cite{mpb+98}), as of late July, includes 2195
bursts. To insure uniform efficiency, we restrict our study to the
1277 bursts triggered on channels 2+3 with a threshold of 5.5 sigma on
all three BATSE time scales. We eliminate from this list 307 GRBs with
peak flux below 0.42 photons $\mbox{cm}^2$ $\mbox{s}^{-1}$, where
BATSE's efficieny is near unity. Efficiency drops radidly below this
cutoff, and uncertainties in efficiency contribute significantly to
estimated number counts (\cite{phm98}). Using a KS test, we compare
the remaining 970 bursts to 5000 GRBs generated from each model.

\begin{figure}
\centerline{\epsfig{file=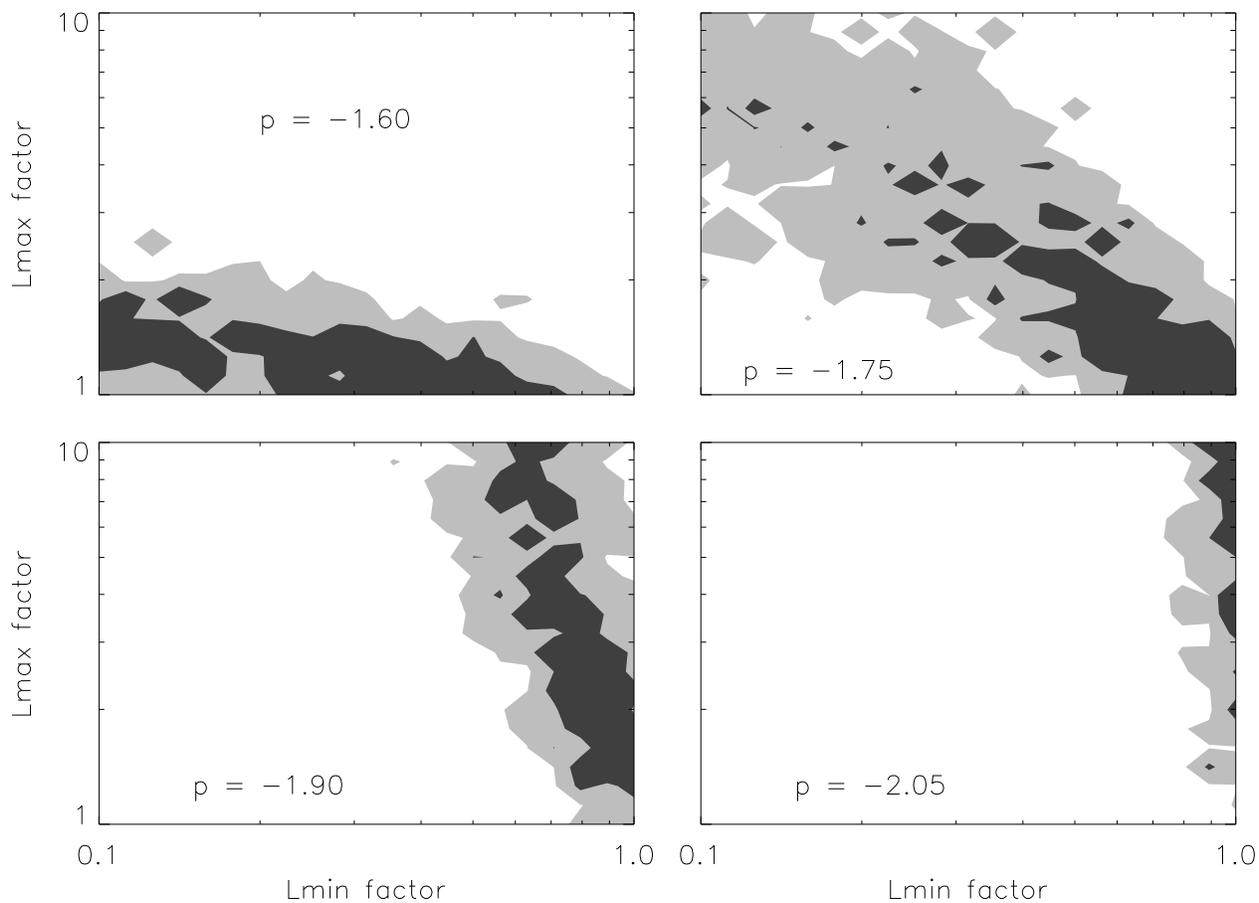}}
\caption{
The light contour shows models with
significance greater than 0.05, and the dark contour shows models with
significance greater than 0.33. One minus the significance value is
the confidence level at which the model can be ruled out.  $L_{\rm
min}$ factor and $L_{\rm max}$ factor indicate the amount by which the
luminosities of GRB970508 and GRB971214 are multiplied,
respectively. Values of $p$, the exponent of the luminosity
distribution, are indicated in each panel.\label{grbmap}}
\end{figure}

The resulting confidence levels are shown in Figure
\ref{grbmap}. The data clearly show there are a number of luminosity
distributions consistent with a redshift distribution that goes as the
star formation history. Our results are consistent with previous work
indicating that the BATSE data constrain most GRBs to a single decade
in luminosity (\cite{hhe+96}).
Despite this constraint, the BATSE data presently allow a very wide
range of possibilities.

\section{Case Study of Two Models}
\label{case}

\begin{figure}
\centerline{\epsfig{file=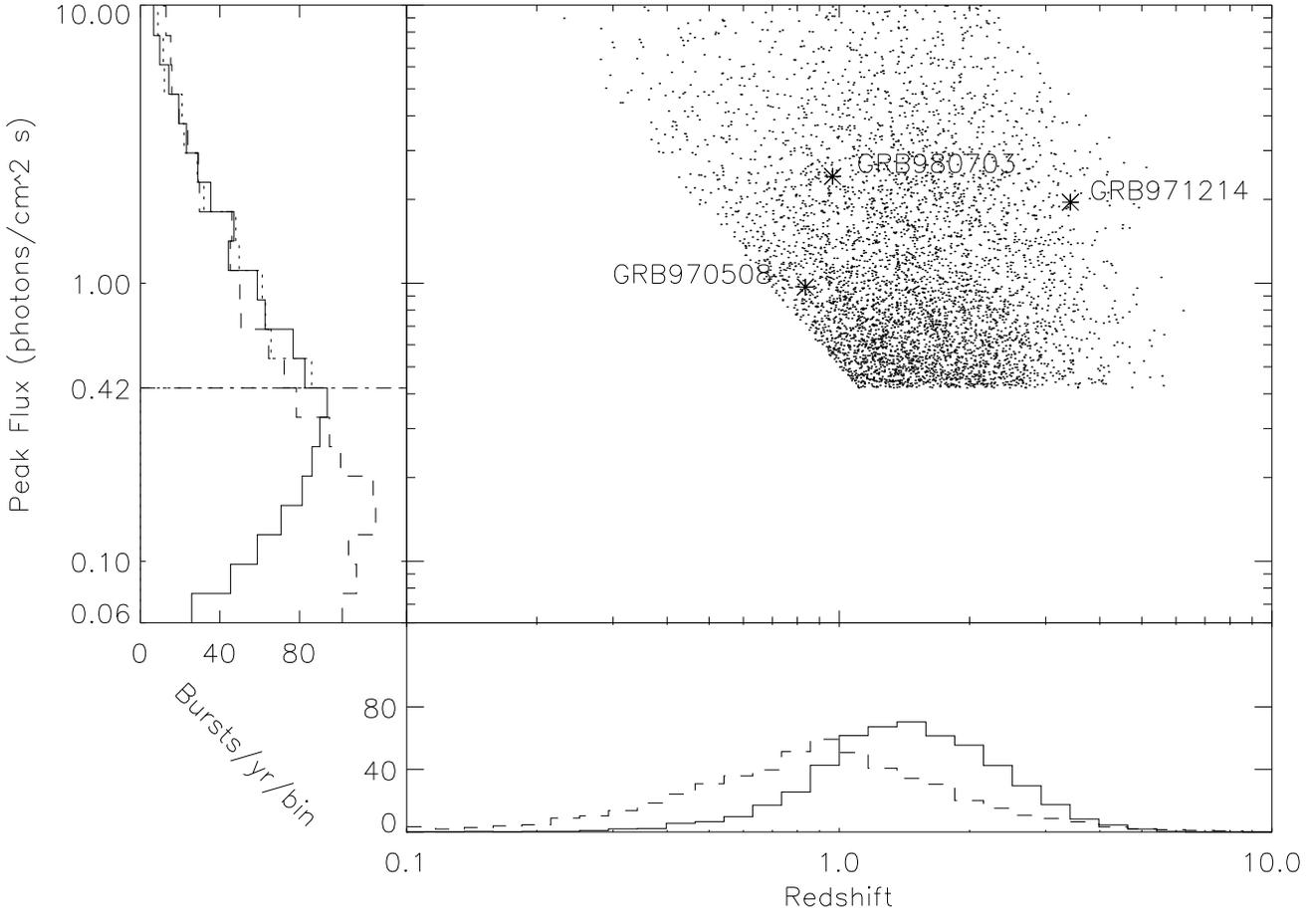}}
\caption{
The center panel shows points indicating the
peak flux and redshift for 5000 GRBs selected from the SFR
distribution with the BATSE flux cutoff; the asterisks are GRB970508,
GRB971214, and GRB980703. The left panel shows the number
of GRBs per year in each peak flux range, divided into 20 evenly spaced
logarithmic bins. The dot-dash vertical line shows the BATSE flux cutoff.  The
histogram in the bottom panel shows the expected number of bursts per
year in each redshift range, divided into 30 evenly spaced logarithmic bins.
The total number of bursts is generated by
setting the number in each model equal to the number per year seen by
BATSE, corrected for BATSE's effective coverage of 48.3 percent of the
sky. In left and bottom panels, the solid histogram is the SFR model,
and the dashed histogram is the CCD model. The dotted line in the left
panel is the BATSE calatog.\label{triplot}}
\end{figure}

To determine what measurements can differentiate the models, we study
two models, which we call the star formation rate model (SFR) and the
constant comoving density model (CCD). SFR uses the star formation
rate as its redshift distribution, and has $L_{\rm min}=1.52\times
10^{50}$ erg/s, $L_{\rm max}=1.16\times 10^{52}$
erg/s, $p=-1.9$ for its luminosity distribution. The CCD model
has a constant comoving redshift distribution, an $L_{\rm
min}=1.13\times 10^{50}$ erg/s, $L_{\rm max}=1.14\times
10^{52}$ erg/s, $p=-2.30$ for its luminosity
distribution. Both these models fit the BATSE data with a KS
significance higher than 0.33, and are also reasonably consistent with
GRB970508, GRB971214, and GRB980703. Figure \ref{triplot} shows
where these three GRBs lie in comparison to the SFR model.

We test whether number versus flux measurements from an instrument
more sensitive than BATSE can differentiate the models. We consider an
instrument with a flux cutoff of 0.06 photons $\mbox{cm}^{-2}$
$\mbox{s}^{-1}$, and perfect efficiency
above this cutoff. Figure \ref{triplot} shows that, while by
construction the models predict identical numbers of bursts below the
BATSE cutoff, they differ substantially at lower peak fluxes. Thus,
the observed number of bursts per year can differentiate the
models. As an interesting side note, in the SFR model, 48 percent of
all GRBs in BATSE's field of view are seen with a flux cutoff of 0.42
photons $\mbox{cm}^{-2}$ $\mbox{s}^{-1}$, and 98 percent are
above 0.06 photons $\mbox{cm}^{-2}$ $\mbox{s}^{-1}$.

We do a KS comparison of the CCD model to a variable number of GRBs
chosen from the SFR distribution. Our results indicate that more
sensitive instruments than BATSE may be able to differentiate our
sample models using only the number-flux relation. One hundred and
twenty-five bursts are enough for a two-sigma differentiation 95
percent of the time. The hypothetical instrument could make this
observation in about 4 months, assuming 50 percent sky coverage and
unit efficiency.

While these results are encouraging, it seems unlikely that
observations of flux alone would do this well in all cases. Based on
$N(<S)$, one model with many bright GRBs at high redshift might well
be completely indistinguishable from another with dim GRBs at low
redshift. In this case, redshift determinations would be necessary to
differentiate the models.

As shown in Figure \ref{triplot}, the two models predict substantially
different redshift distributions for the bursts. In order to determine
the number of redshift measurements required to distinguish the two
models, using BATSE's efficiency, we compared the CCD model to a
variable number of bursts from the SFR distribution in a two-sided,
two-dimensional KS test on the peak fluxes and redshifts.

Our results indicate that about 25 redshifts are needed
to distinguish the SFR and CCD models at the two sigma level 95
percent of the time. Our method is, however, a significant
oversimplification: we assume that it is equally easy to measure any
redshift. In practice, different measurement techniques result in an
efficiency with strong and complicated redshift dependence, with
redshifts between $\sim 1$ and 2, where the bulk of star formation
occurs, difficult to measure from the ground.

Furthermore, two-dimensional KS tests are somewhat unreliable with
small numbers of data points (\cite{ptvf92}). Parameters for the
two-dimensional KS test are derived by numerical experiment, and
assume that the number of samples is large enough for statistical
treatment. Samples with fewer than about 20 data points give results
that depend on the shape of the underlying distribution. Thus, our
figures should be regarded only as lower bounds. Even so, we note that
there are three measured redshifts out of fifteen accurate
localizations. Given this efficiency, we would need more that 125
localizations to reach even this lower limit.

\section{Conclusion}

Our results indicate that the current data available from BATSE are
insufficient to determine the redshift distribution of GRBs. The
complicated mixing between luminosity and comoving redshift
distributions make it impossible to significantly constrain either one
at present. Our example models demonstrate that even a relatively small
change in luminosity distribution can render both a star formation
model and a constant comoving density model consistent with
observations.

The increased flux sensitivity provided by the next generation of
satellites should be able to greatly reduce the parameter space
available. If our comparison of the SFR and the CCD model is
representative, more sensitive satellites may be able to confirm
whether GRBs trace the star formation history even without localizing
any bursts. However, redshift measurements provide considerably
stronger constraints on GRB models.

These results indicate that the best way to constrain the spatial
distribution of GRBs is by directly measuring a sizable number of
redshifts.  As has been demonstrated by the Beppo-SAX localizations,
this is best accomplished by rapid ground-based optical spectroscopic
followup of accurately-positioned bursts, or optical spectroscopy of
the host galaxies. A lower limit on the number of GRB localizations
required to distinguish between disparate models, based on current
efficiency, is more than a hundred.  This is in excess of what will
likely be achieved by continued operation of Beppo-SAX, and by the
upcoming HETE-2 mission (which expects to localize $<\sim 50$ bursts). It
is likely, however, that more rapid notification to the ground of
positions at the arc-minute or better level ($\sim 8'$ SAX positions
currently take 8 hours before distribution, and refinement to $1'$
typically requires a day) will increase the efficiency for redshift
determination by absorption spectroscopy of the optical afterglow. If
this is the case, it is possible that the next generation of
satellites may be able to test whether the spatial distribution of GRB
sources follows star formation.

\acknowledgements{The authors wish to thank S. M. Fall, S. Charlot,
and Y. C. Pei for their data on the cosmic star formation history,
David Hogg for useful discussions, and the referee for helpful
comments on the manuscript. This work is partially supported by a
grant from the National Science Foundation.}

%replace this with the bbl file
%\bibliographystyle{apj1c}
%\bibliography{journals_apj,grbrefs,psrrefs}

\clearpage

\begin{thebibliography}{}

\bibitem[Band {\it et al.}  1993]{bmf+93}
Band, D. {\it et al.}  1993, ApJ, 413, 281

\bibitem[Djorgovski {\it et al.}  1998]{dkg+98}
Djorgovski, S.~G. {\it et al.}  1998, GCN notice 139

\bibitem[Emslie \& Horack 1994]{eh94b}
Emslie, G.~A. \& Horack, J.~M. 1994, ApJ, 435, 16

\bibitem[Fall, Charlot, \& Pei 1996]{fcp96}
Fall, S.~M., Charlot, S., \& Pei, Y.~C. 1996, ApJ, 464, L43

\bibitem[Hakkila {\it et al.}  1996]{hmh+96}
Hakkila, J. {\it et al.}  1996, ApJ, 462, 125

\bibitem[Horack {\it et al.}  1996]{hhe+96}
Horack, J.~M., Hakkila, J., Emslie, A.~G., \& Meedgan, C.~A. 1996, ApJ, 462,
  131

\bibitem[Horv\'{a}th, M\'{e}sz\'{a}ros, \& M\'{e}sz\'{a}ros 1996]{hmm96}
Horv\'{a}th, I., M\'{e}sz\'{a}ros, P., \& M\'{e}sz\'{a}ros, A. 1996, ApJ, 470,
  56

\bibitem[Kippen {\it et al.}  1997]{kwc+97}
Kippen, R.~M. {\it et al.}  1997.
\newblock IAU circular 6789

\bibitem[Kippen {\it et al.}  1998]{k+98}
Kippen, R.~M. {\it et al.}  1998, GCN notice 143

\bibitem[Kulkarni {\it et al.}  1998]{kdr+98}
Kulkarni, S.~R. {\it et al.}  1998, Nature, 393, 35

\bibitem[Loredo \& Wasserman 1998]{lw98}
Loredo, T.~J. \& Wasserman, I.~M. 1998, ApJ, 502, 75

\bibitem[Madau 1997]{mad97}
Madau, P. 1997.
\newblock IAU circular 186

\bibitem[Mallozzi, Pendleton, \& Paciesas 1996]{mpp96}
Mallozzi, R.~S., Pendleton, G.~N., \& Paciesas, W.~S. 1996, ApJ, 471, 636

\bibitem[Meegan {\it et al.}  1998]{mpb+98}
Meegan, C.~A. {\it et al.}  1998, Available at
  http://www.batse.msfc.nasa.gov/data/grb/catalog

\bibitem[Metzger {\it et al.}  1997]{mdk+97}
Metzger, M.~R., Djorgovski, S.~G., Kulkarni, S.~R., Steidel, C.~C., Adelberger,
  K.~L., Frail, D.~A., Costa, E., \& Fronterra, F. 1997, Nature, 387, 879

\bibitem[Paczy\'nski 1986]{pac86}
Paczy\'nski, B. 1986, ApJ, 308, L43

\bibitem[Pendleton, Hakkila, \& Meegan 1998]{phm98}
Pendleton, G.~N., Hakkila, J., \& Meegan, C.~A. 1998, in { 4th Huntsville
  Gamma-Ray Burst Symposium}, ed.\ C.~A. Meegan, R. Preece, \& T. Koshut,
  (Woodbury, New York: AIP), 899

\bibitem[Pian {\it et al.}  1998]{pfb+98}
Pian, E. {\it et al.}  1998, ApJ, 492, L103

\bibitem[Press {\it et al.}  1992]{ptvf92}
Press, W.~H., Teukolsky, S.~A., Vetterling, W.~T., \& Flannery, B.~P. 1992, {
  Numerical Recipes: {T}he Art of Scientific Computing, 2$^{nd}$ edition},
  (Cambridge: Cambridge University Press)

\bibitem[Rutledge, Hui, \& Lewin 1995]{rhl95}
Rutledge, R.~E., Hui, L., \& Lewin, W. H.~G. 1995, MNRAS, 276, 753

\bibitem[Totani 1998]{tot98}
Totani, T. 1998, preprint, astro-ph/9805263

\bibitem[Weinberg 1972]{wei72}
Weinberg, S. 1972, { Gravitation and Cosmology: {P}rinciples and Applications
  of the General Theory of Relativity}, (New York: Wiley)

\bibitem[Woosley 1993]{woo93}
Woosley, S.~E. 1993, ApJ, 405, 273

\end{thebibliography}

\end{document}